\documentclass[lettersize,journal]{IEEEtran}
\usepackage{amsmath,amsfonts}
\usepackage{algorithmic}
\usepackage{algorithm}
\usepackage{array}
\usepackage[caption=false,font=normalsize,labelfont=sf,textfont=sf]{subfig}
\usepackage[most]{tcolorbox}
\usepackage{textcomp}
\usepackage{stfloats}
\usepackage{multicol}
\usepackage{url}
\usepackage{verbatim}
\usepackage{graphicx}
\usepackage{cite}
\usepackage{makecell}
\usepackage[normalem]{ulem}
\usepackage[left=0.673in,right=0.667in,top=0.75in,bottom=1.1in]{geometry}
\usepackage{color}

\begin{document}

\title{Interference Management Strategies for HAPS-Enabled vHetNets in Urban Deployments}

\author{\IEEEauthorblockN{Afsoon Alidadi Shamsabadi\IEEEauthorrefmark{1}, Animesh Yadav\IEEEauthorrefmark{2}, and Halim Yanikomeroglu\IEEEauthorrefmark{1}}\\
\IEEEauthorblockA{\IEEEauthorrefmark{1}Carleton University,  Ottawa, ON, Canada, and \IEEEauthorrefmark{2}Ohio University, Athens, OH, USA}}
\maketitle
\thispagestyle{empty}
\begin{abstract}
Next-generation wireless networks are \mbox{evolving} \mbox{towards} architectures that integrate terrestrial and \mbox{non-terrestrial networks~(NTN)}, unitedly known as vertical \mbox{heterogeneous} \mbox{networks~(vHetNets)}.
This integration is vital to address the increasing demand for coverage, capacity, and new services in urban environments. Among NTN platforms, high altitude platform stations (HAPS) play a promising role in future vHetNets due to their strategic positioning in the stratosphere. In HAPS-enabled vHetNets, various tiers can operate within the same frequency band, creating a \mbox{harmonized-spectrum integrated network}. Although this harmonization significantly enhances spectral efficiency, it also introduces challenges, with interference being a primary concern. This paper investigates vHetNets comprising HAPS and terrestrial macro base stations (MBSs) operating in a shared spectrum, where interference becomes a critical issue. The unique constraints of HAPS-enabled vHetNets further complicate the interference management problem. To address these challenges, we explore various strategies to manage interference in HAPS-enabled vHetNets. Accordingly, we discuss centralized and distributed approaches that leverage tools based on mathematical optimization and artificial intelligence (AI) to solve interference management problems. Preliminary numerical evaluations indicate that distributed approaches achieve spectral efficiency comparable to the centralized algorithm, while requiring lower complexity and less reliance on global information.
\end{abstract}
\section{Introduction}
The sixth-generation (6G) wireless networks aim to revolutionize connectivity by introducing a transformative \textit{beyond connectivity} paradigm, leveraging the advancements achieved in previous generations. This vision goes beyond simply improving the broadband experience, and offers supporting novel use cases and applications that require stringent key performance indicators (KPIs). The International Telecommunication Union (ITU) has outlined the capabilities of International Mobile Telecommunications for 2030 and beyond (IMT-2030), emphasizing key usage scenarios for future networks~\cite{IMT-2030}. 
Primarily, 6G networks are expected to meet the growing demand for immersive communications, hyper-reliable and low latency services, all of which are central to the core scenarios defined by IMT-2030.
Achieving the ambitious objectives of 6G, while ensuring an acceptable quality of experience (QoE) for consumers, necessitates the development and deployment of innovative technologies and network architectures~\cite{6GRoad}.
On the one hand, technologies, such as terahertz communication, ultra massive multiple-input multiple-output (umMIMO) antenna architectures, reconfigurable intelligent surfaces, as well as novel computational techniques based on artificial intelligence (AI) and quantum computing, hold great promise for advancing wireless communications. On the other hand, non-terrestrial networks (NTN) are expected to play a complementary role in the achievement of the 6G goals, specifically enabling the beyond connectivity paradigm~\cite{6GRoad,NTN6G}.

Fundamentally, NTN encompasses space-based satellites and aerial platforms, including high altitude platform stations (HAPS) and uncrewed aerial vehicles. In particular, HAPS refers to quasi-stationary nodes operating in the stratosphere at altitudes between $20$ km and $50$ km, typically moving within a defined cylindrical region~\cite{HAPSSurvey,HAPS-Alouini}. Due to its lower altitude compared to space satellite networks, HAPS presents a promising platform to complement terrestrial networks not only by improving coverage, but also by enhancing spectral and energy efficiency.
While standalone HAPS platforms address diverse use cases in future wireless networks, their integration with terrestrial networks to form vertical heterogeneous networks (vHetNets) unlocks additional advantages.
Specifically, HAPS as IMT base stations, referred to as HIBS by the ITU, can be seamlessly integrated with existing terrestrial networks, forming HAPS-enabled vHetNets. Deploying HAPS-enabled vHetNets in urban regions, with an overlapping coverage between HAPS and terrestrial network, is a potential network architecture, that enables the achievement of ambitious KPIs for 6G while ensuring ubiquitous and reliable connectivity. In the remainder of this paper, the term ``HAPS" refers to ``HIBS".

At the World Radio Conference 2023 (WRC-2023), a new spectrum allocation (sub-2.7 GHz), previously designated for IMT, was extended to include HAPS~\cite{itu_wrc}. This significant development, coupled with the scarcity of available spectrum, underscores the value of a harmonized spectrum approach for vHetNets. By allowing multiple network tiers to operate within the same frequency band, harmonized spectrum vHetNets achieve improved spectral efficiency. However, this configuration is susceptible to performance degradation caused by inter- and intra-tier interference~\cite{our-CL, SpecSharing}. Effective interference management techniques are therefore critical for harmonized spectrum vHetNets. The challenge becomes even more complex due to the altitude and other unique constraints associated with HAPS. As a result, interference management solutions designed for standalone terrestrial networks cannot be applied directly to vHetNets; they must be redesigned with appropriate objectives and new constraints specific to HAPS.

To address these issues, this paper presents the following contributions:
\begin{itemize} 
    \item First, the distinct characteristics of HAPS-enabled vHetNets are identified and discussed.
    \item Second, several potential interference management strategies, including conventional techniques adapted to include HAPS-specific constraints and newly proposed methods, are suggested for implementation in HAPS-enabled vHetNets to enhance overall
performance.
    \item Third, considering the complexities of the interference management formulated problems, an overview of efficient mathematical optimization and AI-driven tools, to solve interference management problems in vHetNets, is provided.
    \item Finally, the advantages of distributed algorithms over centralized approaches are demonstrated through numerical results.
\end{itemize}
\section{Unique Characteristics of HAPS-enabled vHetNets}\label{Sec:vHetNet_Interference}
\begin{figure}[t]
    \centering
    \captionsetup{justification=centering}
    \includegraphics[width=\linewidth]{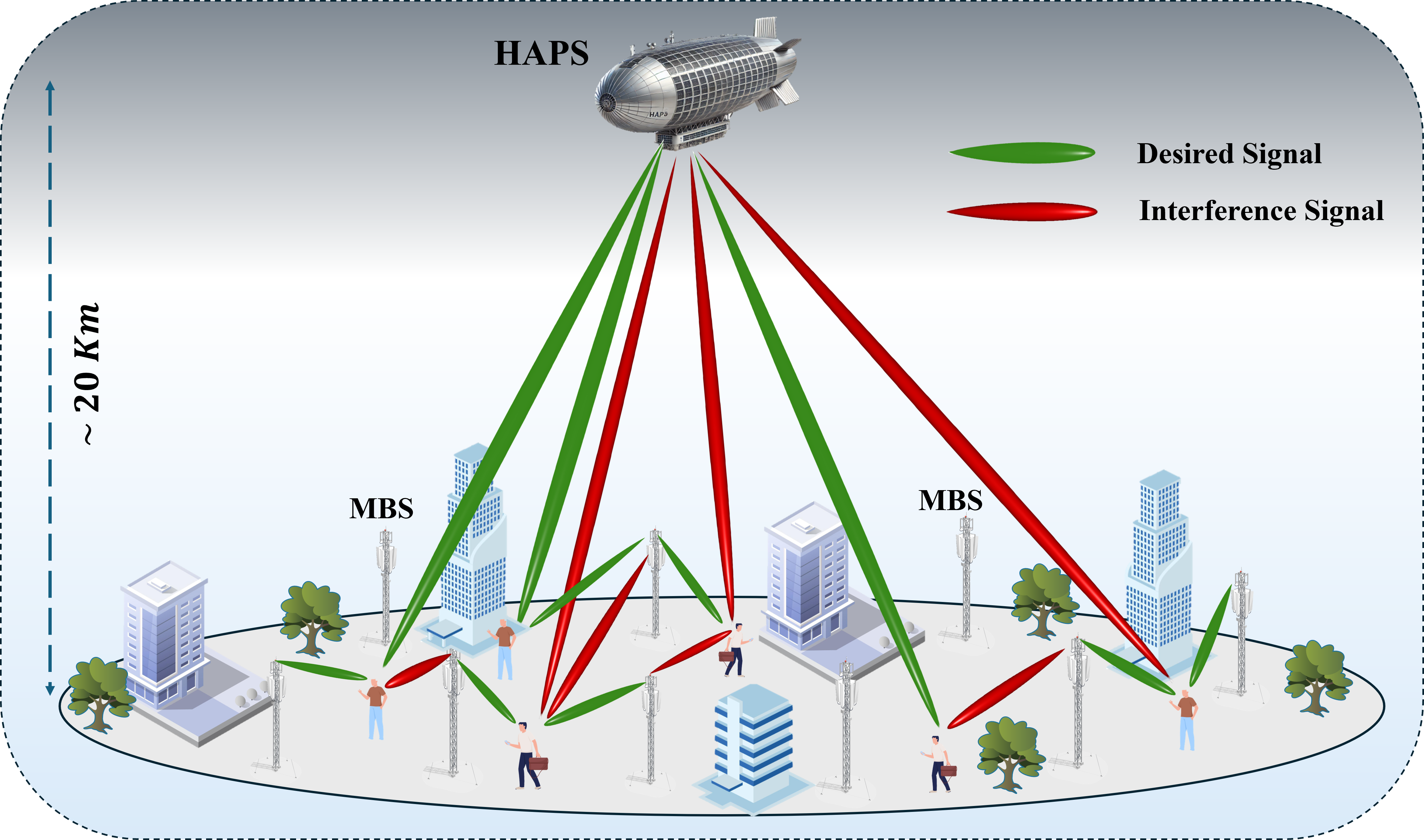}
    \caption{System model of an integrated HAPS-terrestrial network (vHetNet).}
    \label{fig_1}
\end{figure}
HAPS-enabled vHetNets are a promising architecture to enhance user experience in dense urban regions. However, adding another tier to the network introduces new constraints and challenges that must be addressed for optimal performance. Moreover, HAPS specific constraints become aggravating in urban environments due to overlapping coverage with terrestrial networks and a higher number of users demanding continuous data traffic.
Therefore, the interference management problem in vHetNets differs from that of terrestrial networks due to distinct conditions and constraints associated with HAPS. These differences render traditional interference management strategies in terrestrial networks \mbox{ineffective} when applied to vHetNets. The key differences are discussed below.
\subsection{Extensive Coverage Area of HAPS}
One of the defining characteristics of vHetNets is their large-scale nature. The high altitude of HAPS allows vHetNets to cover an extensive area, as shown in Fig.~\ref{fig_1}, which encompasses multiple terrestrial macro base stations (MBSs). Therefore, effective coordination between HAPS and MBSs is crucial to ensure optimal service delivery to users. Unlike terrestrial MBSs, which only interact with their neighbors, HAPS must collaborate with all MBSs within its coverage area. This interaction leads to large-scale optimization problems with increased complexity, which in most cases are challenging to solve with rapid convergence. In addition, efficient decision making requires the exchange of information about the status of the network between MBS and HAPS, which adds a huge data overhead to communication links.
\vspace{-0.2cm}
\subsection{HAPS Limited Onboard Energy}
HAPS is uniquely positioned to take advantage of solar energy irrespective of the weather conditions, thanks to its large surface area and above the clouds placement, making it a sustainable option for prolonged operations. However, despite these advantages, the large size and weight of HAPS, coupled with its movement within a cylindrical area, impose significant energy demands. This is further intensified by the complexity of the large antenna architectures and the sophisticated payloads, which require a substantial amount of power to function effectively. These constraints are especially important when dealing with interference management problems in HAPS deployments, as energy efficiency becomes a crucial factor. 
This underscores the need for advanced energy-aware interference management techniques in HAPS systems to maximize their operational efficiency and sustainability.
\subsection{HAPS Limited Wireless Backhaul Link Capacity}
Unlike MBSs, which use high-capacity fiber links for core network connectivity, HAPS relies on wireless links for connection to ground stations and then to the core network via fiber links. However, these wireless links are constrained by limited capacity, imposing challenges in terms of user association and, hence, the network scalability. The limited capacity of these wireless links restricts the number of users that can be efficiently served by a HAPS at any given time. Consequently, this constraint must be carefully considered when designing interference management algorithms for HAPS-enabled vHetNets, ensuring optimal, robust, and scalable resource allocation schemes and guaranteed QoE for the associated users.
\subsection{HAPS Channel Model and Impact on Objective Function} 
With a high probability, HAPS establishes a strong line-of-sight (LoS) connection with users due to strategic positioning in the stratospheric layer. Even in urban areas where shadowing and non-LoS (NLoS) links are prominent, the users experience strong LoS connections from HAPS along with small-scale fading channels. This robust LoS connection, coupled with low-variance free-space path loss (FSPL), compensates for shadowing and NLoS effects. As a result, users generally experience consistent channel quality with HAPS. However, when HAPS is integrated with terrestrial networks, the strong LoS connection can cause most users to be associated with HAPS, which is not feasible due to its limitations. This scenario necessitates the design of appropriate objective functions for optimization problems. Additionally, HAPS' LoS connection with users leads to the spatial correlation between the channels of different users with HAPS. Therefore, the signal received by the users will be correlated.
\subsection{HAPS Jittering}
 Due to its positioning, HAPS introduces a certain degree of jitter into the communication link with ground-based users. Jitter can be caused by wind or the circular flight paths of HAPS over a region. This can lead to instability in communication channels and disrupt the synchronization between HAPS and terrestrial MBSs. When users compete for the same resources, these disruptions significantly complicate coordination efforts, leading to degraded network performance and unreliable data transmission.

The aforementioned constraints and characteristics highlight the complexities involved in developing effective interference management algorithms for vHetNets. To maximize the benefits of vHetNets, a rethink of the interference management strategy is essential. In the next section, some potential strategies for managing interference in vHetNets are discussed.
\section{Interference Management Strategies in HAPS-enabled vHetNets}\label{Sec:Approaches}
\begin{figure}[t]
    \centering
    \captionsetup{justification=centering}
    \includegraphics[width=\linewidth]{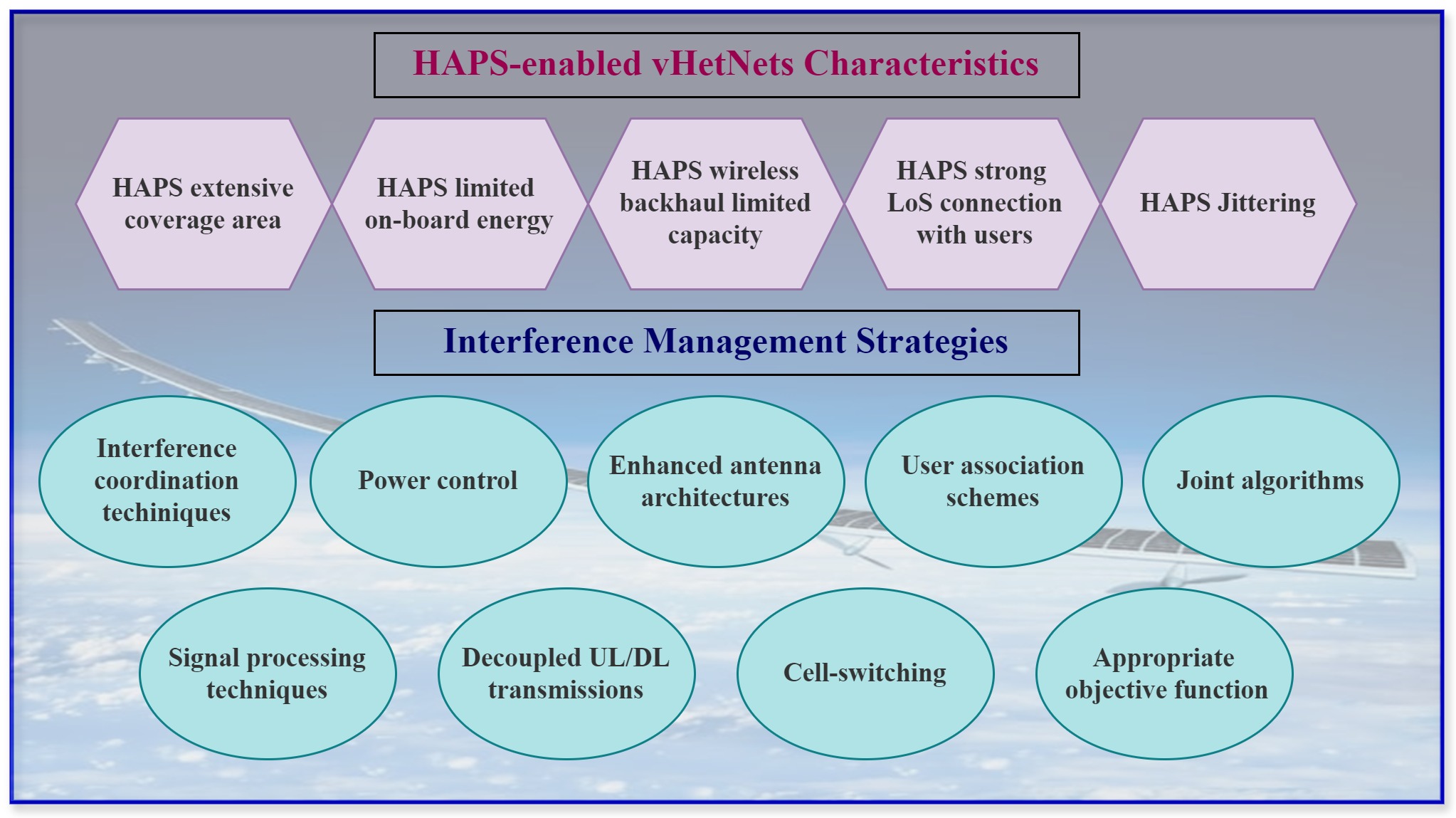}
    \caption{Interference management in HAPS-enabled vHetNets: Distinct characteristics and potential strategies.}
    \label{fig2: Challenges and Approaches}
\end{figure}
Interference management has long been a fundamental challenge in wireless networks, and related strategies have continuously evolved to keep pace with the complexities of emerging network architectures. Different approaches have been proposed across various layers of wireless networks to mitigate interference. In the following, we discuss some potential strategies applicable to HAPS-enabled vHetNets, as summarized in Fig.~\ref{fig2: Challenges and Approaches}.
\subsection{Interference Coordination Techniques}
Interference coordination (IC) techniques have traditionally been proposed for use in wireless networks, encompassing both static and dynamic schemes such as frequency reuse and coordinated multipoint (CoMP) techniques. Although the increasing demand for higher data rates often limits the effectiveness of these algorithms, they remain promising solutions for managing interference.
In particular, within large-scale vHetNets, the overall coverage area can be divided into clusters, where IC strategies can be employed to reduce inter-cluster interference. Addressing intra-cluster interference continues to be a significant challenge in these scenarios, however, with reduced complexity.
\subsection{Power Control}
In vHetNets, ground users experience a relatively consistent FSPL when communicating with HAPS. This is due to the high altitude of the HAPS, which results in a near constant distance from the ground over large areas, reducing the variance of FSPL experienced by different users. Therefore, the need for dynamic adjustment of transmission power becomes critical to ensure that all users receive adequate signal strength without affecting other users. Effective power control can optimize spectral efficiency for each user, minimize interference, and conserve \mbox{energy}~\cite{Our-WCL}. 
\subsection{Enhanced Antenna Architectures and Beamforming}
Enhanced umMIMO antenna architectures significantly improve the performance of wireless networks by forming user-specific narrow beams. The beam width and gain of the main lobe and side lobes play an important role in reducing interference. The large surface area of HAPS presents a unique opportunity to deploy umMIMO antennas with a substantial number of antenna elements. HAPS can generate pencil beams toward users by dynamically allocating antenna elements and utilizing enhanced beamforming techniques. However, optimal beamforming designs such as zero-forcing approaches are not applicable to large-scale vHetNets due to the high computational costs. Therefore, efficient and rapid converging algorithms for robust beamforming weights should be designed~\cite{our-CL}. In addition to the umMIMO and beamforming approaches, advanced arrangements of the umMIMO antenna elements (such as cylindrical and hemisphere antennas) are another potential approach for interference management. These advanced antenna arrays facilitate creating beams with less spatial correlation, and therefore less correlation between the received signals, at the same time with increasing the degree of freedom in managing interference.
\subsection{User Association Scheme}
Another effective strategy to manage interference is the design of proper user association schemes. In vHetNets, given the large coverage area of HAPS and the uniqueness of the channel between HAPS and users, associating users who are in close proximity to each other with HAPS can lead to significant interference, even with appropriate beamforming design. This is because the relatively short distances between users become negligible compared to the altitude of the HAPS. Therefore, an efficient association of users with base stations\footnote{The term ``base station" refers to both HAPS and terrestrial MBSs.} is essential to manage and minimize the propagated interference~\cite{our-ICC}. Further, by relaxing the constraint on single base station association, associating users with multiple base stations including HAPS and MBSs is another promising scheme to improve the performance of vHetNets~\cite{our-CL}.
\subsection{Joint Algorithms}
Although various standalone approaches can mitigate interference in vHetNets, combining these methods further enhances algorithm performance and provides additional gains. For instance, a joint user association and beamforming weight design algorithm yields better results compared to standalone user association or beamforming algorithms \cite{our-CL}. Similarly, a joint subcarrier allocation and power control algorithm delivers improved performance compared to the standalone approaches~\cite{Our-WCL}.
\subsection{Signal Processing Techniques for Residual Interference}
Despite the aforementioned efficient approaches for mitigating and managing interference in large-scale vHetNets, interference cannot be completely eliminated, especially due to the jittering effect in HAPS. In particular, within HAPS-enabled vHetNets, the HAPS jitter results in a lack of synchronization between the signals received from HAPS and MBSs. Therefore, jitter-aware advanced signal processing techniques are required at the transmitter and receiver to analyze the structure of the interference to suppress or cancel any residual interference.
\subsection{Decoupled UL/DL Transmissions}
A completely different approach to mitigate interference, specifically for HAPS-enabled vHetNets, is to decouple the uplink (UL) and downlink (DL) transmissions between the MBSs and HAPS. In particular, for a frequency division duplex (FDD) transmission, where UL and DL links operate on different frequency bands, when the users associated with MBSs transmit in the UL channels, the users associated with HAPS transmit in the DL channels, and vice versa.
\subsection{Cell-Switching Strategy}
Cell-switching is also a potential approach to optimize energy efficiency and decrease interference in wireless networks. Particularly, cell-switching can be considered as an interference management technique for multi-tier networks, such as HAPS-enabled vHetNets. In cell-switching scenarios, some MBSs can completely offload their users to HAPS and turn off their transmission in order to reduce interference and improve the energy efficiency of the network~\cite{CellSwitching}. Therefore, developing cell-switching algorithms with the multi objective of energy efficiency enhancement and interference management is a promising energy-aware interference management strategy.
\subsection{Appropriate Objective Function}
Designing an appropriate objective function for interference management in HAPS-enabled vHetNets is a critical consideration. In standalone HAPS networks, where the LoS connection is dominant, the weighted sum rate is an effective objective function. However, in vHetNets, where users may experience a poor channel with terrestrial networks but a strong LoS connection with HAPS, proportional fairness serves as the optimal objective function, as it balances both fairness and the total sum rate of the network~\cite{our-ICC}. Therefore, interference management strategies for vHetNets must ensure proportional fairness among users. 

The aforementioned approaches typically involve formulating constrained optimization problems and solving them to find optimal solutions. However, under HAPS-specific constraints, the formulated problems are challenging to solve. In the following section, we discuss some potential methodologies that can help solve these optimization problems in vHetNets, aiming for suboptimal yet practical and rapid converging solutions.
\section{Interference Management: Efficient Mathematical Tools}
\begin{figure*}[t]
\centering
\captionsetup{justification=centering}
\includegraphics[width=0.9\linewidth]{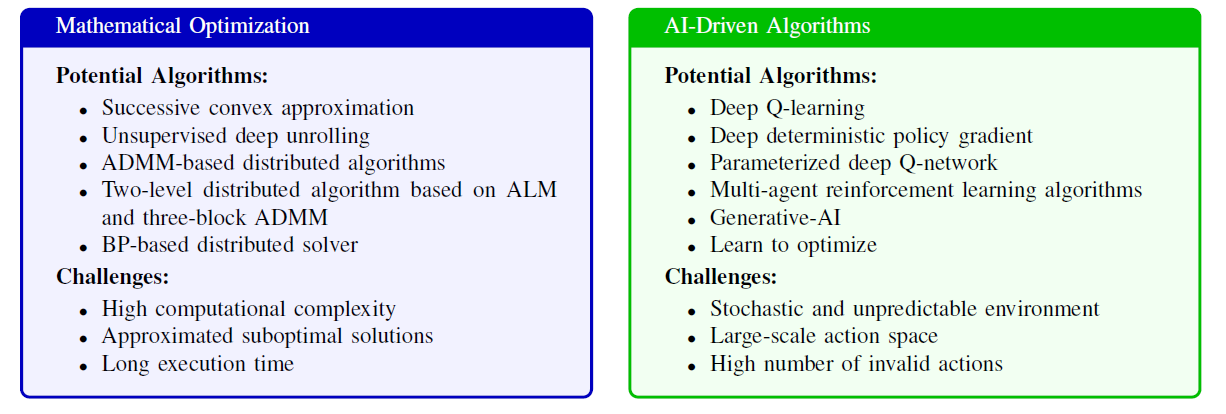}
\caption{Interference management in vHetNets: Efficient methodologies.}\label{tab:table1}
\end{figure*}

In the previous section, we provided some potential strategies that can be employed in vHetNets to manage or mitigate interference. The optimization problems for interference management in HAPS-enabled vHetNets deal with strictly non-convex objective function and constraints. For example, using a logarithmic utility function for proportional fairness objective function introduces the logarithm of a logarithm in the objective function, which adds non-convexity to the problem. Additionally, constraints such as the limited wireless backhaul link capacity of HAPS further add to the non-convexity. Given the complexity of these formulated optimization problems, finding an optimal solution is a difficult task. Therefore, novel and efficient tools must be developed to find approximate solutions. In the following of this section, we provide an overview of some potential methodologies, also summarized in Fig.~\ref{tab:table1}, to address these challenges.

\subsection{Centralized and Distributed Mathematical Optimization \mbox{Algorithms}}
Given the non-convexity and intractable nature of the interference management problems associated with HAPS-enabled vHetNets, various equivalent transformations, approximations, and relaxations must be applied to convert the original problem to a tractable and convex equivalent format. These techniques, collectively known as reformulation and linearization techniques (RLT), include methods such as first-order Taylor series linearization, exponential cone approximations, and other mathematical approximations that provide linear or convex lower or upper bounds for non-convex functions. The RLT converts the original problem into approximate but convex problems, which can then be solved iteratively, widely known as the successive convex approximation (SCA). At the end of the SCA process, the optimal solution to the approximated problem is a suboptimal solution to the original problem. Although SCA provides the suboptimal solution with rapid convergence, iteratively solving multiple optimization problems in a centralized manner presents several challenges, as outlined below.
\begin{itemize}
    \item In centralized algorithms, network status information, including channel coefficients, must be sent to a central node. In large-scale vHetNets, HAPS can serve as the central node, collecting channel coefficients from all MBSs within its coverage area. However, this leads to increased latency and significant communication overhead between HAPS and MBSs. 
    \item Centralized algorithms must solve the optimization problem to obtain all network variables, greatly increasing the complexity of the problem, making centralized solutions impractical for large-scale vHetNets.
\end{itemize}

To overcome these shortages, the approaches based on deep unrolling have the potential to solve the constrained non-convex problems. Such approaches, such as unsupervised deep unrolling methods, which rely on projection gradient descent, are promising tools, as a replacement for iterative approaches, to solve non-convex constrained optimization problems~\cite{Unfolding}. 
Alternatively, developing distributed algorithms offers an effective solution by solving the optimization problem locally at each base station. Distributed approaches provide the following advantages compared to centralized algorithms:
\begin{itemize}
    \item Enhanced scalability due to reduced information exchange among base stations while still achieving the global network objective.
    \item Lower computational complexity, as each base station solves a smaller local optimization problem rather than a large centralized one.
    \item Increased resilience, as distributed algorithms do not rely on a centralized node, eliminating potential single points of failure."
\end{itemize}
To this end, several distributed approaches, well suited for problems in wireless systems, have been proposed in the literature, with one of the most common being the alternating direction method of multipliers (ADMM), designed to solve convex optimization problems. However, since the objective functions and constraints in vHetNets are non-convex, the problem must first be converted into the SCA form, after which ADMM is applied in each SCA iteration. Although this approach reduces data overhead between base stations, combining SCA and ADMM at each base station increases the execution time of the algorithm. To address this and to develop a distributed algorithm for non-convex problems without the need for SCA, a potential solution is to employ the two-level distributed algorithm based on an augmented Lagrangian method (ALM) with an inner loop of the three-block ADMM approach~\cite{TWoStepDistributed}. This approach extends beyond traditional ADMM by introducing key reformulations into the original non-convex problem, making it suitable for a three-block ADMM algorithm. In this way, the distributed algorithm can be effectively applied to design interference management algorithms in large-scale vHetNets.

Another potential approach to developing a distributed algorithm in vHetNets is the belief propagation (BP) methodology, which aims to find optimal solutions at each local node~\cite{BPMIMO}. Although BP-based methods do not require a central node,
they are not well-suited for dense, large-scale vHetNets because each local node must compute all possible solutions to find the optimal one. 
To address this, an efficient approach to decrease the complexity is to first approximate the original problem using linear programming, and then solve the approximated problem using a BP-based distributed solver~\cite{BPMIMO}.

\subsection{AI-Driven algorithms}
In recent years, AI-enabled wireless networks have attracted considerable interest from researchers. AI and machine learning technologies play a pivotal role in facilitating the integration of terrestrial networks with NTN. Moreover, AI-driven algorithms show significant potential for managing interference in vHetNets. Interference management schemes often involve making sequential decisions about the system parameters. In this context, various reinforcement learning (RL)-based algorithms have emerged as promising solutions. Trained RL agents can determine optimal or near-optimal parameters through a one-shot decision-making process, thereby reducing the complexity and execution time compared to traditional optimization-based algorithms.
Depending on the characteristics of the action and state spaces, several RL algorithms, such as deep Q-learning, deep deterministic policy gradient (DDPG), and parameterized deep Q-network (PDQN), can be employed. However, RL-based algorithms face significant challenges that hinder their scalability and feasibility in large-scale vHetNets, as discussed in the following.
\begin{itemize}
    \item  \textit{Agent training in random environments: }RL algorithms train agents to make decisions based on the state of the environment. In vHetNets, key parameters that influence interference, such as channel coefficients, are tied to the environment, making them suitable candidates for the input of neural networks. However, these channel coefficients are random variables, which makes it challenging to train agents effectively in such dynamic environments. Many existing approaches simplify this problem by using fixed channel coefficients during training that significantly limits the real-world applicability of the trained agent to handle dynamic and stochastic environments.
    \item \textit{Continuous vs. discrete action spaces:} Most RL algorithms are designed for discrete action spaces, whereas parameters like power allocation and beamforming weights in vHetNets are continuous. Although power variables can be discretized into predefined levels, this strategy does not extend well to beamforming weights, which require continuous representation. Addressing continuous and hybrid action spaces can be achieved using algorithms, such as DDPG and PDQN, although these come with added complexities.
    \item \textit{High-dimensional action spaces:} The large number of variables in vHetNets results in high-dimensional action spaces for RL algorithms. This increases training time and often hinders the agent's ability to learn optimal actions across the entire space. Potential solutions include adopting alternative policies capable of managing large-scale networks or leveraging multi-agent RL algorithms~\cite{LargeScaleRL}.
    \item \textit{Constraints and feasibility of actions:} The constraints discussed in Section \ref{Sec:vHetNet_Interference}, such as limited transmit power and HAPS wireless backhaul capabilities, render many potential actions invalid. Although user-specific constraints can be managed using suitable activation functions at the neural network output layer, global constraints, such as the total transmit power allocated to all HAPS users, pose a greater challenge in vHetNets.
\end{itemize}

Considering the above challenges, designing enhanced AI agents which are capable to deal with dynamic environments of wireless networks and take hybrid actions under certain constraints is an open challenge.
In addition to conventional AI algorithms, Generative-AI (GenAI) is also a promising tool for designing the physical layer algorithms in wireless networks. Furthermore, GenAI can be used to generate training data for RL agents, making them more robust in making decisions. Furthermore, leveraging AI, traditional optimization methods can be enhanced to be combined with AI algorithms. In this context, learn-to-optimize (L2O) is designed to enhance performance through training, which usually occurs online. At the same time, this training aims to make the actual application of the methods more efficient. In situations involving complex problems, such as nonconvex optimization or inverse problems, a well-trained L2O method might deliver better results than traditional methods.
In the following section, we provide a comparison between centralized and distributed optimization algorithms, highlighting the advantages of distributed algorithms when dealing with large-scale vHetNets.
\section{Case Study}
In this case study, we consider a cell-free vHetNet consisting of $1$ HAPS and $4$ MBSs, simultaneously serving $U$ single antenna users in a rectangular urban area measuring $4$ km by $4$ km. We assume that HAPS is equipped with $8 \times 8$ antennas, while MBSs are equipped with $4 \times 4$ antennas. In addition, all users are assumed to share the same time-frequency resources.
We use a Rician channel with a Rician factor of $10$ to model the channel between HAPS and users, which accounts for a strong LoS connection alongside small-scale fading due to the urban environment. For the channel between MBSs and users, we consider small-scale fading, shadowing, and FSPL.
The system operates in a harmonized spectrum within the sub-$6~\text{GHz}$ frequency band, with a center frequency of $2.545~\text{GHz}$. The total transmit power available for HAPS and MBSs is assumed to be $52~\text{dBm}$ and $43~\text{dBm}$, respectively.
To efficiently manage interference in this network, we implement the beamforming weight design (BWD) algorithm in both centralized and distributed approaches, aiming to maximize proportional fairness under total available power, HAPS wireless backhaul link, and minimum signal-to-interference-plus-noise ratio constraints. The centralized algorithm is implemented according to \cite{our-ICC}, and we implement a distributed algorithm based on the two-level distributed approach.

\begin{figure}[t]
    \centering
    \captionsetup{justification=centering}
    \includegraphics[width=\columnwidth]{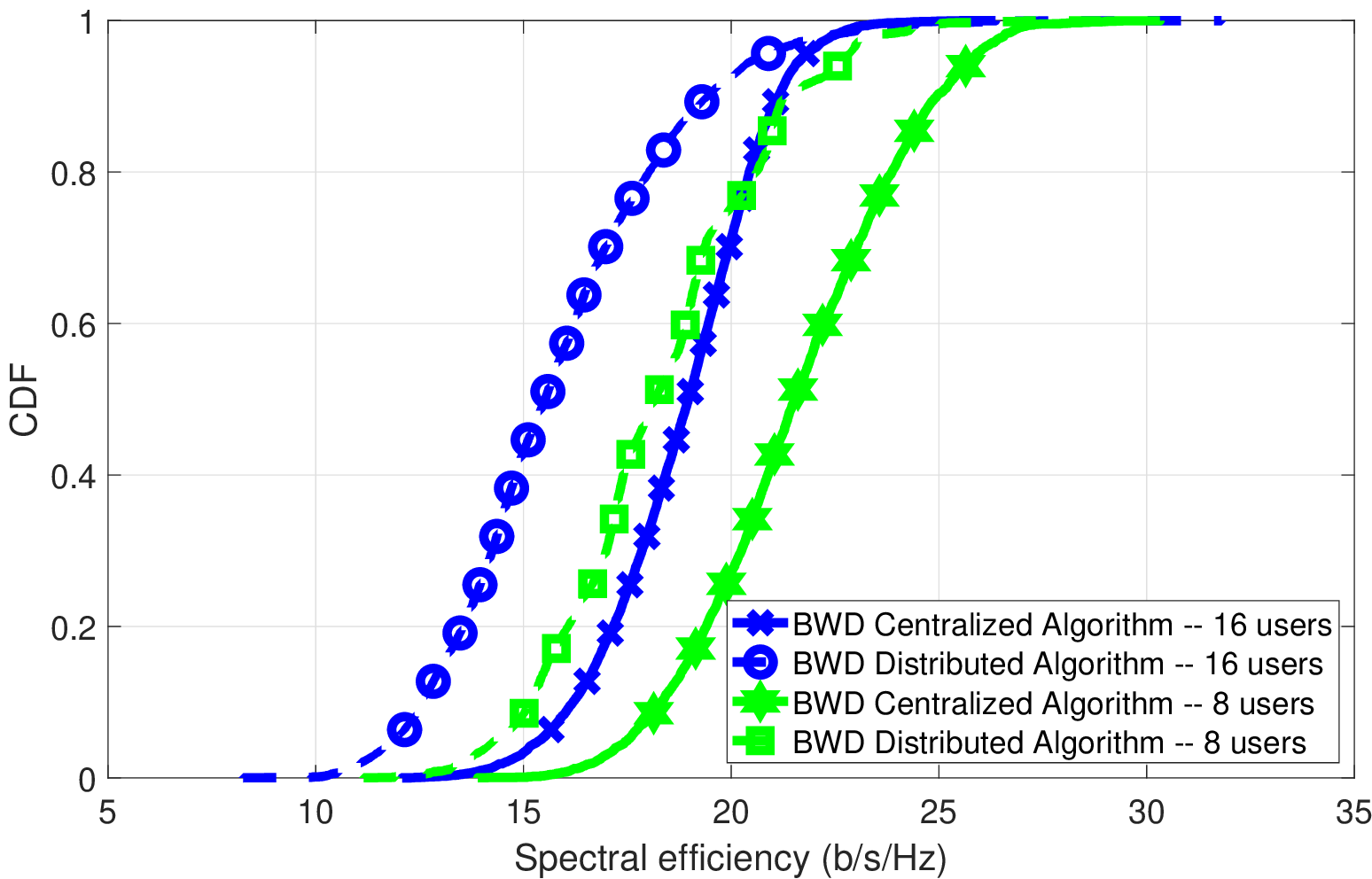}\label{fig_4:b}
    \caption{Comparison between distributed and centralized mathematical optimization in a HAPS-enabled vHetNet.}
    \label{fig_4}
\end{figure}
Figure~\ref{fig_4} shows the cumulative distribution function (CDF) of spectral efficiency for both centralized and distributed BWD algorithms. The algorithms were tested for $U=\{8,16\}$ users. From the comparison of the algorithms' performance, three key observations can be made. First, with a significant reduction in complexity, the performance of the distributed algorithm in terms of the CDF of spectral efficiency is comparable to that of the centralized algorithm, and the decrease in the achieved performance is not significant. Second, the distributed algorithm significantly reduces the complexity of the optimization problem in terms of the number of variables to be solved at each base station, making it applicable to dense large-scale vHetNets. Finally, solving the centralized problem requires the central node to collect all channel coefficients from the base stations, which introduces substantial overhead in the information exchange among the base stations. These findings lay the foundation for future research, which will further develop and benchmark advanced interference management algorithms, incorporating detailed comparisons with existing approaches.
\section{Conclusion}
In this work, we discussed the integration of HAPS with the terrestrial networks as a promising multi-tier network architecture for 6G, referred to as HAPS-enabled vHetNet. We discussed harmonized spectrum vHetNets and the importance of addressing the interference challenge in vHetNets. Consequently, we provide an overview of the challenges and constraints that are unique to HAPS and make interference management problems in vHetNets different from standalone terrestrial networks. Accordingly, some potential strategies to manage interference in vHetNets were proposed. Finally, a few AI-driven and optimization-based potential algorithms were outlined, aligned with their challenges. Through preliminary simulation results of the case study, we validated the efficiency of using distributed algorithms over centralized interference management algorithms to handle large-scale vHetNets.
\bibliographystyle{IEEEtran}
\bibliography{ref.bib}
\section*{Biography}
\vspace{-34pt}
\begin{IEEEbiographynophoto}{Afsoon Alidadi Shamsabadi}
    [SM] (afsoonalidadishamsa@sce.carleton.ca) is a PhD Candidate in the Carleton-NTN (Non-Terrestrial Networks) Lab of the Systems and Computer Engineering Department at Carleton University, Canada. Her research focuses on the innovative integration of high-altitude platform stations (HAPS) with terrestrial wireless networks, and AI-enabled wireless networks.
\end{IEEEbiographynophoto}
 \vspace{-34pt}
\begin{IEEEbiographynophoto}{Animesh Yadav}
    [SM] (yadava@ohio.edu) is an assistant professor in the School of EECS at Ohio University, Athens, OH, USA. His research interest includes intelligent communications, computing and sensing for 6G and beyond networks. He is an IEEE Senior Member and currently serving as a Senior Editor for IEEE Communications Letters and Associate Editor for Frontiers in Communications and Networks journals.
\end{IEEEbiographynophoto}
\vspace{-34pt}
\begin{IEEEbiographynophoto}{Halim Yanikomeroglu} [F] (halim@sce.carleton.ca) is a Chancellor’s Professor in the Department of Systems and Computer Engineering at Carleton University, Canada, and he is the Director of Carleton-NTN (Non-Terrestrial Networks) Lab. His group’s focus is the wireless access architecture for the 2030s and 2040s, and non-terrestrial networks. He is a Fellow of EIC (Engineering Institute of Canada), CAE (Canadian Academy of Engineering), and AAIA (Asia-Pacific Artificial Intelligence Association).
\end{IEEEbiographynophoto}

\vfill

\end{document}